\documentstyle[sprocl]{article}

\input{psfig.tex}

\def\Journal#1#2#3#4{{#1} {\bf #2}, #3 (#4)}

\def\PLB{{\em Phys. Lett.}  B}
\def\PRL{\em Phys. Rev. Lett.}
\def\PRD{{\em Phys. Rev.} D}

\def\be{\begin{equation}}
\def\ee{\end{equation}}
\def\bea{\begin{eqnarray}}
\def\eea{\end{eqnarray}}

\begin{document}

\title{HADRONIC ELECTROMAGNETIC FORM FACTORS AND COLOR TRANSPARENCY }

\author{BIJOY KUNDU, PANKAJ JAIN}

\address{ Department of Physics, IIT Kanpur, Kanpur-208 016, India }

\author{JOHN P. RALSTON}

\address{
Department of Physics and Astronomy, University of Kansas, Lawrence,
KS 66045, USA }

\author{JIM SAMUELSSON} \address{Department of Theoretical Physics,
Lund University, Lund, Sweden}


\maketitle\abstracts{ We review the current status of electromagnetic
form factor calculations in perturbative QCD. There is growing
evidence that factorization prescriptions involving a transverse
coordinate integration, such as that of Li and Sterman, are more
appropriate than the prescription of LePage and Brodsky.  Color
transparency is naturally described within the formalism.  We report
the first explicit calculations of color transparency and nuclear
filtering as perturbatively calculable phenomena.}


\section{Introduction}
The applicability of perturbative QCD to exclusive processes
has always been controversial.  Despite the remarkable agreement with
data of the quark-counting scaling laws of Brodsky and Farrar, the
helicity conservation selection rules of Lepage and Brodsky tend not
to agree with data \cite{GRF,GPL,CZ}.

It has not been clear how to interpret this conflict.  By
dimensional analysis, scaling indicates that a finite, minimal number
of quarks is being probed.  However, the failure of hadronic helicity
conservation is a direct test of the factorization scheme, and the
failure cannot be repaired by appealing to models of distribution
amplitudes or their normalizations.  Failure of hadron helicity
conservation apparently rules out dominance by the short distance
formalism.  It has been common to identify the short-distance
formulation as being ``the same as'' perturbative QCD ($pQCD$) itself.  Then
the agreement of the scaling laws with experiments appears to be
rather mysterious.

Theoretical criticisms focus on calculations found to include regions
where the internal momentum transfers are too small for leading order
$pQCD$ to apply reliably \cite{Isg,Rad}.  For even the simplest model
calculations, the case of hadronic form factors, it is found that
large contributions come from the components of quark wave functions
involving large quark spatial separations.

A reasonable resolution of the conflict observes that a factorization
scheme is merely a tool, in which different amplitudes are
re-arranged for the purpose of calculation.  Hence if one factorization
scheme is inapplicable to experiment, one can always try another,
and the underlying applicability of the approximations may be
improved.  Li and Sterman \cite{LS,L} gave an improved factorization
formula for calculating the pion and the proton form factor, which
included Sudakov suppression.  The Sudakov form factor tends to
suppress the regions of large quark spatial separations, thereby
extending the applicability of $pQCD$. We will review the mechanism and
our calculations in detail below.

Unfortunately the Sudakov effect is not sufficiently dramatic to
resolve all the issues in the region of current momentum transfers
($Q^{2}$).  This provides an additional strong motivation for
extending the experimental scope, and in particular for studying
quasi-exclusive reactions in nuclear targets.

The early conception of color transparency \cite{brodMuell} was based
on having large momentum transfer $Q^{2}$ select short distance
regions, freed with color coherence to propagate through a passive nuclear
probe.  Given
the controversies over short-distance dominance, it has not been clear
whether large enough $Q^{2}$ would be obtained to make the basic
assumptions apply at laboratory energies. However, large quark separations
should tend to be absorbed in the
strongly interacting nuclear medium, while small quark separations
should penetrate freely\cite{JB,RP90,PBJ}.  This is the phenomenon
called ``nuclear filtering'', which acts somewhat like Sudakov
suppression.  Instead of $Q^{2}$ as the large dimensionful scale,
there is the large nuclear radius of order $A^{1/3} fm$.

Both Sudakov effects and nuclear
filtering depend directly on the transverse coordinate.  In fact, the
transverse-position space factorization in which color transparency
and nuclear filtering is described \cite{RP90} pre-dates the very similar
factorization of Li and
Sterman \cite{LS,L}.  Both hearken to the proton-proton
scattering work of Botts and Sterman \cite{BS}, which was
constructed to address the inability of Lepage-Brodsky factorization
to describe independent scattering.  Because all of the ideas spring from a
common
factorization prescription, the explicit calculations dovetail
together perfectly, and they can be presented in the same format.

\section{Hadronic Form Factors}

LePage and Brodsky \cite{GPL} calculate the pion electromagnetic form
factor at momentum transfer $q^{2} = -Q^{2}$ with a factorization
method written as
\begin{equation}
F_\pi(Q^2) = \int dx_1 dx_2 \phi(x_2,Q) H(x_1,x_2,Q) \phi(x_1,Q).
\end{equation}
Here $\phi(x,Q) $ are the distribution amplitudes, which can be
expressed in terms of the pion wave function $\psi(x,\vec k_T)$ as
\begin{equation}
 \phi(x,Q) = \int^Q d^2k_T\psi(x,\vec k_T) .
\end{equation}
We use $x$ for the longitudinal momentum fraction and $\vec k_T$ for
the transverse momentum carried by the quark.  The factorization is
justified provided the external photon momentum $Q^2$ is
{\it asymptotically large}.  Then the $k_T$ integrals decouple, and the
$k_T$ dependence of the hard scattering $H$ can be expanded in a power
series, retaining the trivial, constant term.  One directly obtains the
power-law
scaling of the quark-counting method, with logarithmic corrections.

Unfortunately, several authors \cite{Isg,Rad} have found that much of
the numerical weight of explicit calculations comes from the end point
regions.  The proposed decoupling of the transverse integrations is
not a numerically accurate approximation, making application of
the method suspect.  We reiterate that this legitimate source of doubt
has often been extended to the whole application of $pQCD$
\cite{Isg,Rad} to hard exclusive processes.  Since $pQCD$ and the
factorization scheme are separate concepts, it is unjustified to jump
to such a conclusion.

The Li-Sterman factorization \cite{LS,BS} retains coupling of the
$k_T$ dependence of the wave functions and the hard scattering.  In
some sense the concept is less ambitious theoretically, by including a
broader integration region than the zero-distance LePage-Brodsky
method. The calculation is simplified by dropping the weak
$k_T$ dependence of quark propagators in a hard scattering kernel $H$.
No loss of consistency occurs, because the rest of the $k_{T}$
dependence is sufficient to justify this step.  Working in
configuration space \cite{LS} the usual convolutions become a product:
\begin{equation}
F_\pi(Q^2) = \int dx_1 dx_2 {d^2\vec b\over (2\pi)^2} {\cal
P}(x_2,b,P_2,\mu) \tilde H(x_1,x_2,Q^2,\vec b,\mu) {\cal
P}(x_1,b,P_1,\mu),
\end{equation}
where ${\cal P}(x,b,P,\mu)$ and $\tilde H(x_1,x_2,Q^2,\vec b,\mu)$ are
the Fourier transforms of the wave function, including Sudakov
factors, and hard scattering respectively; $\vec b$ is conjugate to
$\vec k_{T1} - \vec k_{T2}$, $\mu$ is the renormalization scale and
$P_1$, $P_2$ are the initial and final momenta of the pion.

The leading-order Li-Sterman method is marginally consistent in
practice.  If the
dependence on a transverse separation cutoff \cite{LS} is studied at
$Q^{2}$ of a few $GeV^{2}$,
then nearly 50\% of the form factor comes from a region where
$\alpha_s/\pi < 0.7$.  This indicates that higher order contributions
in $\alpha_s$ may not be negligible, but this is a separate issue from
the factorization scheme.  Certainly the leading order predictions for
the normalization of the form factor cannot be regarded as accurate.
Next to leading order calculation \cite{passek} of the pion form
factor confirm this conclusion.  Numerically, the fact that results of
the calculations (described in detail elsewhere \cite{Jain}) lie
below the experimental data cannot be given great weight; in fact, the
agreement is actually quite acceptable.

\subsection{The Proton}
The proton Dirac form factor $F^p_1(Q^2)$\cite{L} is considerably more
complicated.  In contrast to the pion, there is no natural choice for the
infrared cutoff in the Sudakov exponent, due to the presence of three
quarks and resulting three distances.

Bolz et. al \cite{JKB} pointed out that the infra-red
cut-off $b_c$ used by Li \cite{L} does not suppress the soft
divergences as $b_c \rightarrow 1/\Lambda_{QCD}$. A modified choice
of the cutoffs was proposed by them \cite{JKB}.  Subsequently the
form factor was found to saturate as $b_c \rightarrow
1/\Lambda_{QCD}$.  The normalization of the resulting $Q^4F_1$ was
found to be less than half of that of the data for all the
distribution amplitudes explored \cite{JKB}.  Bolz et al \cite{JKB}
then concluded that $pQCD$ is unable to fit the experimental form
factor.

Kundu et al \cite{KLSJ} re-examined the situation.  Considerably more
complete calculations were performed\cite{KLSJ}, incorporating the
full two-loop correction to the Sudakov effects.  A physical choice of
the infra-red cutoff parameter was also incorporated.  This cut-off
prescription treats the proton as a quark-diquark
configuration at the extreme point of quark separation.

As a result, Kundu et al \cite{KLSJ} find that the
calculation is in good agreement with data using the King-Sachrajda
(KS) \cite{KS} distribution amplitude.  The fact that the
normalization of the proton form factor can be fit makes an important
conceptual point: the method in principle can explain the data, if
higher order corrections in $\alpha_{s}$ were under control.  Again
this is supported by examination of the contributing integration
regions, or $b_c$ dependence \cite{KLSJ}.  Saturation occurs at about
$b_c = 0.8/\Lambda_{QCD}$, so about 50\% of the calculation comes from
the soft or the large $b$ regions.  Scaling is postponed to beyond
$Q^2=10 GeV^2$, but is inevitable after that.

The pion and the proton form factor calculations reveal that
short-distance regions, required to be dominant in the basic
LePage-Brodsky factorization, do not dominate in practice.  Other
prescriptions (such as Li and Sterman) are capable of incorporating
long-distance regions.  The $Q^2$ scaling dependence of the proton
form factor above about $10
GeV^2$ appears to be quite robust. The calculations are sufficiently
independent of
the theoretical uncertainties such as distribution amplitude models,
the infra-red cut-off parameters, and higher-order corrections, to
indicate that current large $Q^2$ experimental scaling is truly
fundamental.

\section{Color Transparency and Nuclear Filtering}

Color transparency is a natural
prediction of $pQCD$. However, if the asymptotic limit is taken prematurely
(as
in LePage-Brodsky factorization) then all targets have perfect
transparency, and there is nothing left to calculate. Taking $Q^{2}$
indefinitely large (but fixed), one might think all targets
become transparent. But then taking $A \rightarrow \infty$ all
targets become opaque.  Thus there is a limit interchange problem in
the LePage-Brodsky factorization, because the limit of large $Q^{2}$
and large $A$ do not commute. The scheme is limited to asymptotic
$Q^{2}$, and fundamentally unable to describe the
phenomena at laboratory energies.

A correct description of the phenomenon follows from a
factorization scheme incorporating the transverse degrees of freedom
\cite{RP90}.  It is very useful that we do not have to rely on
extremely
large $Q^{2}$ to motivate $pQCD$, but instead large $A$ serves as an
infrared cutoff.  The ``filtering limit'' takes $A>>1$ with $Q^{2}$
fixed and large enough to motivate a $pQCD$ approach to attenuation:
this requires merely $Q^{2} >$ a few $GeV^{2}$.  On this basis it has
been predicted that perturbative QCD calculations are more reliable in
a nuclear target.

These concepts have experimental support.  Experimentally one
finds that the fixed-angle free space process $pp'\rightarrow p''p'''$
\cite{Car} shows significant oscillations at 90 degrees as a function
of energy.  The energy region of oscillations extends over the whole
range of high energy measurements that exist, from $s=6 GeV^{2}$ to
$s=40 GeV^{2}$.  The oscillations are large, making up roughly 50\% of
the $1/s^{10}$ behavior, and are interpreted as coming from
interference of long- and short-distance amplitudes.  The corresponding
process in a nuclear environment $pA\rightarrow p'p''(A-1)$ shows no
oscillations, and obeys the $pQCD$ scaling power law far better than the
free-space data\cite{JB,BT88,PBJ}.  The $A$ dependence, when analyzed
at fixed $Q^2$, shows statistically significant evidence of reduced
attenuation\cite{JainRalPRD}.

While the formalism and model calculations have existed for a while,
the calculations to verify it are quite complex.  Only with the
completion of the work by Li et. al \cite{LS} were all the pieces to make the
complete perturbative calculation laid out in ordered form.  The
calculations require Monte Carlo integrations of very high order (up
to 9 dimensions after taking into account all symmetries) which have
never before been attempted.  The results, however, are encouraging
and show that the nuclear interactions do substantially eliminate the
soft region \cite{KJ}.
\begin{figure} [t] \hbox{\hspace{6em}
\hbox{\psfig{figure=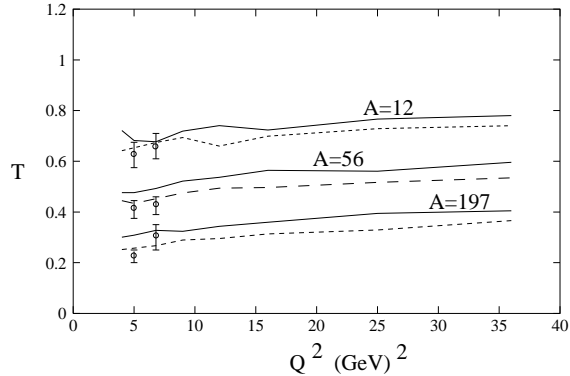,height=2in}}} \caption{The calculated
transparency ratio for the proton for different nuclei.  The experimental
points are taken from Ref. [29,30].
The solid curves are calculated with $k=5$ and the
dashed curves with $k=6$.} \label{fig1}
\end{figure}

\begin{figure} [t] \hbox{\hspace{6em}
\hbox{\psfig{figure=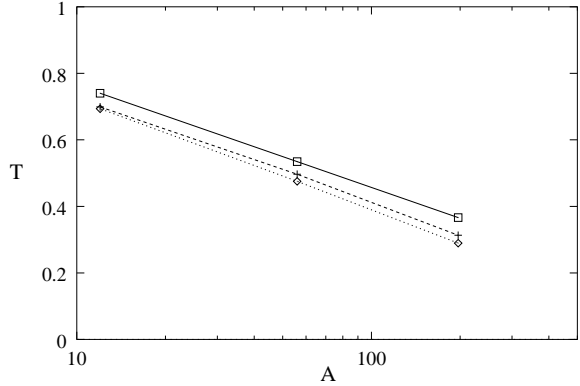,height=2in}}} \caption{The calculated
transparency ratio for the proton as a function of $A$ for different $Q^2$. The
solid, dashed and dotted curves correspond to $Q^2=36,\ 16$ and 9 GeV$^2$
respectively. The value of the parameter $k=6$.} \label{fig2} \end{figure}

\begin{figure} [t] \hbox{\hspace{6em}
\hbox{\psfig{figure=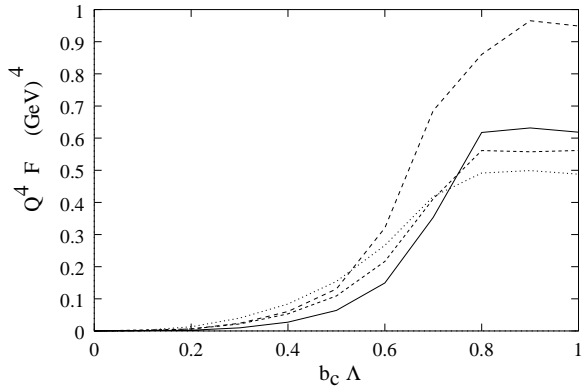,height=2in}}} \caption{The transverse
separation cutoff $b_c$ dependence of the proton amplitude ratio.
Curves are drawn for $A=197$. The solid, short dashed and the dotted
curves are calculated for $Q^2=$ $6.8$, $16$ and $36$ GeV$^2$ respectively. The
long dashed curve corresponds to the free space calculation for $Q^2 = 16 $
GeV$^2$, which contains
substantially more long-distance contamination.} \label{fig3} \end{figure}

\begin{figure} [t] \hbox{\hspace{6em}
\hbox{\psfig{figure=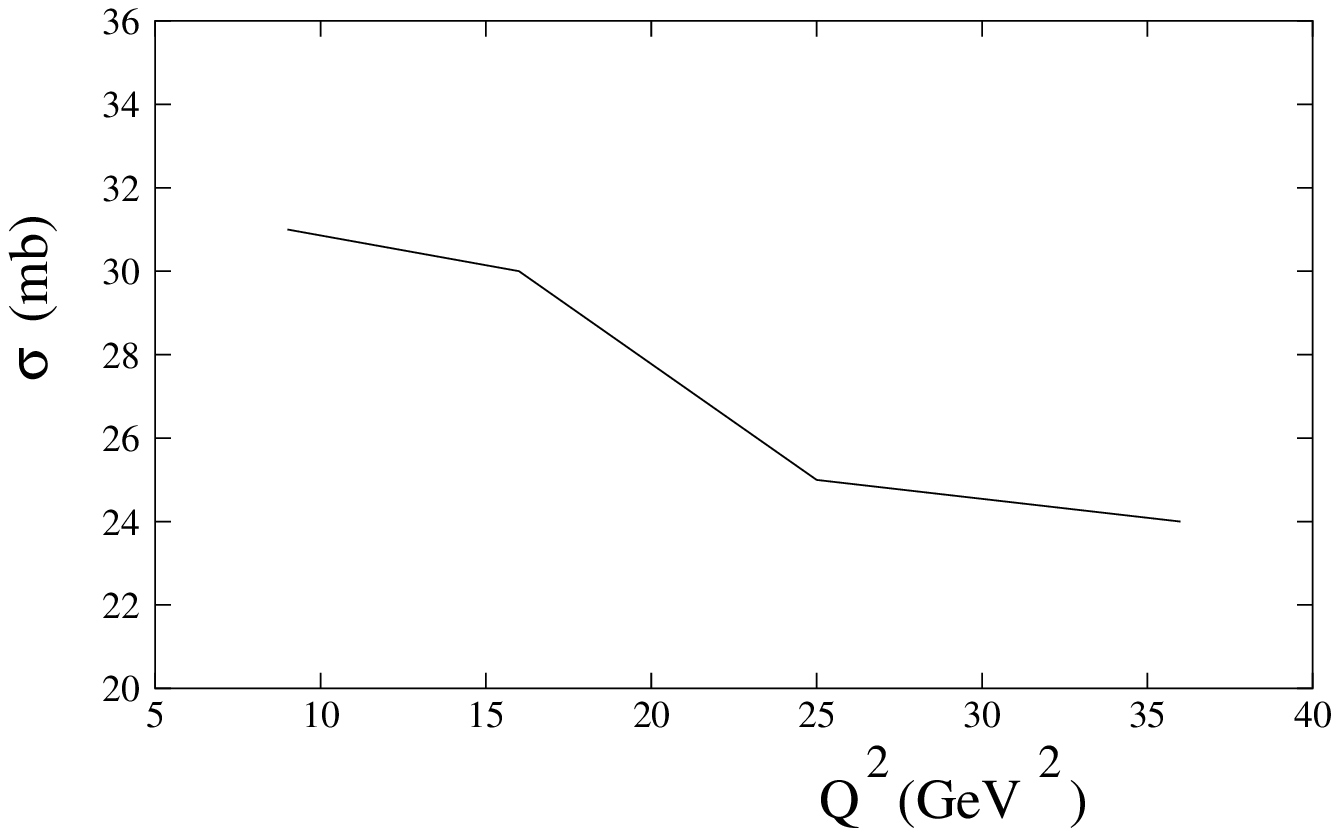,height=2in}}}
\bigskip
\caption{
Extracted effective attenuation cross sections $\sigma_{eff} (Q^2)$ as a
function of $Q^2$ exhibit color transparency.  The calculations fit the
curvature of the $A$ dependence
using the same model of nuclear structure and correlations as
other calculations.  The decrease of $\sigma_{eff} (Q^2)$ with $Q^2$ is
sufficiently large that conventional nuclear physics might be ruled out
with sufficiently large $Q^2$ or sufficiently precise experimental data.
} \label{fig4} \end{figure}

\subsection{\it The Pion: Nuclear Medium Effects}

The nuclear medium modifies the quark wave function such that
\cite{RP90} \begin{eqnarray} {\cal P}_A(x,b,P,\mu) = f_A(b; B){\cal
P}(x,b,P,\mu), \end{eqnarray} where ${\cal P}_A$ is the wave function
inside the medium and $f_A$ is the nuclear filtering amplitude.  An
eikonal form \cite{GL,durand}
appropriate for $f_A$ is:

\begin{eqnarray} f_A(b; B) = exp(-\int_{z}^{\infty} dz' \sigma(b)
\rho(B, z')/2) .
\end{eqnarray} Here $\rho(B, z')$ is the nuclear number density
\cite{Vries} at
longitudinal distance $z$ and impact parameter $B$ relative to the nuclear
center.
We have used the fact that the imaginary part of the eikonal amplitude
for forward scattering is related to the total cross section,
explaining our use of the symbol $\sigma(b)/2$.  Finally, we must
include the probability to find a pion at position $B, z$ inside the
nucleus, which we take to be a constant times the probability to find
a nucleon.  Putting together the factors, the process of knocking out
a pion inside a nuclear target has an amplitude $M$ given by

\begin{eqnarray}
M &=& \int_{0}^{\infty} d^2 B \int_{-\infty}^{+\infty} dz \rho(B,z)
\times F_{\pi}(x_1,x_2,b,Q^2) \times f_A(b,B)
\end{eqnarray} The inelastic cross section $\sigma$ is known to scale like
$b^2$ in
QCD \cite{lonus,gunsop}.  We parametrize $\sigma(b)$ as $k b^2$ and adjust
the value of $k$ to
find a reasonable fit to the experimental data.

\subsection{\it The Proton: Nuclear Medium Effects}

For the proton the important transverse scale is the maximum of the
three quark separation distances, $b_{max} = max(b_1,b_2,b_3)$.
The calculation of the process in the nuclear target needs a 9
dimensional integration, which is performed by Monte Carlo.  The
effects of short-range correlations were included approximately by
replacing \cite{LM} \begin{eqnarray} \rho(z',b)\rightarrow
\rho(z',b)C(|z-z'|), \end{eqnarray} where $C(u)$ is a correlation
function estimated in \cite{MS} to be $C(u) = [g(u)]^{1/2}$ with
\begin{eqnarray} g(u) = \left[1-{h(u)^2\over 4}\right] [1+f(u)]^2
\end{eqnarray} where \begin{eqnarray} h(u) = 3{j_1(k_Fu)\over k_Fu}\ ,
\end{eqnarray} \begin{eqnarray} f(u) = -e^{-\alpha u^2}(1-\beta u^2)
\end{eqnarray} with $\alpha=1.1$, $\beta=0.68$ fm$^{-2}$ and the Fermi
momentum $k_F=1.36$ fm$^{-1}$.

\section {Results and Discussions}
Results for the $Q^{2}$ dependence of the proton transparency ratio are
given in Fig. \ref{fig1}
and \ref{fig2}. The parameter $k$ in the attenuation cross section
$\sigma=kb^2$
was chosen so as to provide a reasonable fit to the experimental data
\cite{Mak,Neill}.

The quark transverse separation cutoff  $b_c$ dependence of the amplitude
ratio is shown in Fig. \ref{fig3}.
It is clear from this figure that the
large distance contributions are significantly reduced in the nuclear
environment. We find that for a heavy nucleus at 36 GeV$^2$, 90\% of
the contribution comes from a region where $\alpha_s/\pi$ is less than 0.7.

We have also checked the dependence of our result on the infrared cutoff
parameter $c$ and the choice of the wave function.
We find that the results for transparency ratio change very little if we
use the CZ
wave
function instead of the KS.
This merits further study.
The result shows some dependence on the parameter $c$, but this dependence
is significantly reduced compared to the case of the free form factor 
\cite{KLSJ}.

Finally, following \cite{JainRalPRD}, we have extracted the effective
attenuation
cross section
$\sigma_{eff} (Q^2)$, which serve as a litmus test of whether ``color
transparency" has actually been achieved.
Our calculations of $\sigma_{eff} (Q^2)$
were done using the same model of correlations and nuclear density as the
rest of our calculations.  The results (Fig. \ref{fig4})
show a significant decrease of $\sigma_{eff} (Q^2)$ with increasing $Q^2$
to values well below the Glauber model attenuation cross section, which
indicates color transparency.

\section *{Acknowledgments}
PJ and BK thank the staff of ICTP, Trieste, for hospitality during a
visit where this paper was written. This work was supported by
BRNS grant No. DAE/PHY/96152,
the Crafoord Foundation, the Helge Ax:son Johnson Foundation, DOE Grant number
DE-FGO2-98ER41079,
the KU General Research Fund and NSF-K*STAR
Program under the Kansas Institute for Theoretical and Computational
Science.

\section *{References} 
\end{document}